\documentstyle[amssymb,amstex,11pt]{article}

\newtheorem{remark}{Remark}[section]
\newcommand{\p}{\partial}
\newcommand{\half}{\frac{1}{2}}
\title{{\bf Classical Theta Functions \\
and Quantum Tori}}
\author{Alan Weinstein\thanks{Research partially supported by NSF
Grant DMS-93-09653.}
\\Department of Mathematics\\
University of California\\
Berkeley, CA 94720 USA\\
{\small(alanw@@math.berkeley.edu)}}
\begin{document}
\maketitle
\begin{abstract}
The Schwartz kernel of the multiplication operation on a quantum torus
is shown to be the distributional boundary value of a classical
multivariate theta function.  The kernel satisfies a Schr\"odinger
equation in which the role of time is played by the deformation
parameter $\hbar$ and the role of the hamiltonian by a Poisson structure.
At least in some special cases, the kernel can be written as a sum of
products of single-variable theta functions.
\end{abstract}

A {\em quantum torus} is, by definition, a ``space'' whose
(noncommutative) algebra of functions is obtained by deformation of
the (commutative) algebra of functions on a torus $({\Bbb
R}/{\Bbb Z})^d$.  In this note, we observe that the Schwartz
kernels of the multiplication operators for a class of quantum tori
are the (distributional) boundary values of classical theta
functions, and we draw some conclusions from this observation.  In
particular, we show that each of these kernels satisfies a
Schr\"odinger equation, in which the role of time is played by the
deformation parameter $\hbar$, and the role of the hamiltonian is
played by the Poisson structure associated with the deformation.
Thus, the ``evolution in Planck's constant'' of a torus is like the
time evolution of a free particle, with the initial state (the
Schwartz kernel of the operator of pointwise multiplication) being a
delta function, like the initial state of a particle with certain
position and totally uncertain momentum.

This evolution equation suggests a possible answer to the question
of how a deformed product must behave after the deformation parameter
$\hbar$ has left the first infinitesimal neighborhood of zero.

We note that our work seems to be quite different from Manin's study
\cite{ma:quantized} of {\em quantum} theta functions, which are special
functions on the quantum tori themselves.

\section{The kernel of multiplication on a quantum torus}
\label{sec-one}

The noncommutative multiplications on the space of functions on ${\Bbb
T}^d$ which we will discuss in this paper are those which define the
so called ``non-commutative tori'', or ``quantum tori'', an important
class of examples of noncommutative differentiable manifolds, surveyed
in \cite{ri:tori}.  Specifically, we consider the
``star products''
$*_{\hbar}$ given in terms of the basis $$\{e_m(x) = e^{2\pi imx}\}_{m \in
{\Bbb Z}^d}$$ of functions on ${\Bbb T}^d$ and the deformation
parameter $\hbar$ by
\begin{equation}
\label{eqn1}
e_m *_{\hbar} e_n = e^{-\pi i\hbar P(m,n)}e_{m+n}
\end{equation}
where $P: {\Bbb R}^{m^*} \times {\Bbb R}^{m^*} \rightarrow {\Bbb R}$
is a skew-symmetric bilinear form.

Notice that $e_m *_0 e_n = e_{m+n}$, which is the rule for ordinary
(commutative) multiplication on ${\Bbb T}^d$, and
\begin{eqnarray*}
\frac {1}{i\hbar} (e_m *_{\hbar} e_n - e_n *_{\hbar} e_m) &= &\frac
{1}{i \hbar} (e^{-\pi i \hbar P(m,n)} - e^{\pi i\hbar P(m,n)})e_{m+n} \\
&= &-2\pi P(m,n)e_{m+n} + O(\hbar),
\end{eqnarray*}
so the Poisson structure which is the semiclassical limit of this
deformation is given by $$\{e_m,e_n\} =-2\pi P(m,n)e_{m+n},$$
or, equivalently, $$\{f,g\}(x) = (1/2\pi )P(df(x),dg(x)).$$

The product defined by (\ref{eqn1}) is initially defined on the space
$C^{\infty}({\Bbb T}^d)$.  The resulting algebra is denoted by ${\cal
A}_{\hbar P}^{\infty}$ or by $C^{\infty}({\Bbb T}_{\hbar P}^d)$, where
${\Bbb T}_{\hbar P}^d$ denotes the ``quantum manifold'' on which
${\cal A}_{\hbar P}^{\infty}$ is the algebra of ``smooth functions''.

\begin{remark}
\em The algebra  $C({\Bbb T}_{\hbar P}^d)$ of {\em continuous} functions on
${\Bbb T}_{\hbar P}^d$ is the $C^*$ algebra obtained by completing
${\cal A}_{\hbar P}^{\infty}$ with respect to a certain norm
\cite{ri:tori}.
There seems to be no natural
identification of the Banach space $C({\Bbb T}_{\hbar P}^d)$ with the
space $C({\Bbb T}^d)$ of continuous functions on the ordinary torus.
As Rieffel points out in \cite{ri:tori}, the elements of $C({\Bbb T}_{\hbar
P}^d)$ are determined by their Fourier coefficients, but the set of
possible Fourier series for these elements depends on $P$ and, just as
in the case $P=0$, is difficult to describe in a simple way.
The results in this paper may make it possible to explain the
difference between $C({\Bbb T}_{\hbar P}^d)$ and
$C({\Bbb T}^d)$ in terms of the singular nature of the distribution
kernel for multiplication in $C^{\infty}({\Bbb T}_{\hbar P}^d)$.
\end{remark}

The product $*_{\hbar}$ for ${\Bbb T}^d$ may also be considered as the
restriction to $C^{\infty}({\Bbb T}^d)$ of the Moyal product
\[
(f *_{\hbar} g) (x)\sim \sum_{j=0}^{\infty} \frac {1}{j!}
\left.\left[ \frac{i\hbar}{4\pi }P \left( \frac {\p}{\p y}, \frac {\p}{\p
z}\right)\right]^{j}(f(y)g(z))\right| _{y=z=x}
f(y)g(z)
\]
on $C^{\infty}({\Bbb R}^d)$ (defining the quantum affine space ${\Bbb
R}_{\hbar P}^d$), where $C^{\infty}({\Bbb T}^d)$ is identified with
the space of $C^{\infty}$ functions on ${\Bbb R}^d$ invariant under
translation by the lattice ${\Bbb Z}^d$.

When $P$ is nonsingular, the ``inverse'' of $P/2\pi $ is a symplectic
structure on ${\Bbb T}^d$ which we will denote by $\omega$.  In this
case, the product
$*_{\hbar}$ is also given by the integral formula
\begin{equation}
\label{eqn2}
(f *_{\hbar} g)(x) = \int_{{\Bbb R}^d \times {\Bbb R}^d} \frac
{1}{(\pi \hbar)^d} e^{2i\omega (x-y,x-z)/\hbar }f(y)g(z)dydz.
\end{equation}
The improper integral in (\ref{eqn2}) can be defined as an oscillatory
integral when $f$ and $g$ lie in the space of functions whose
partial derivatives of all orders are each uniformly bounded, and the
product belongs to the same space.  (See \cite{ho:weyl} for a proof of
the closure of this and other spaces under the Moyal product.) This fact
and translation invariance allow the Moyal product
to pass to $C^{\infty}({\Bbb T}^d)$.

For singular $P$, the Moyal product can still be defined by an
integral formula, but the kernel $K(x,y,z) =
\frac {1}{(\pi \hbar)^d}
e^{2i\omega (x-y,x-z)/\hbar }$ is no longer a smooth function.
Instead, it is
a delta-distribution supported by the set of triples $(x,y,z)$ for
which the differences $x-y$ and $x-z$ (and hence $y-z$ as well) lie in
the range of the map ${\tilde P}: ({\Bbb R}^d)^* \rightarrow {\Bbb
R}^d$ associated with the bilinear form $P$, i.e. when $x$, $y$, and
$z$ all lie in the same symplectic leaf of the Poisson structure $P$.
We will see shortly that the product for ${\Bbb T}_{\hbar P}^d$ is
also given by a distribution kernel, which is singular even if $P$ is
nondegenerate.

To find the kernel of multiplication on ${\Bbb T}_{\hbar P}^d$, i.e.,
the distribution $K_{\hbar P}(x,y,z)$ for which
\[
(f *_{\hbar} g)(x) = \int_{{\Bbb T}^d \times {\Bbb T}^d} K_{\hbar
P}(x,y,z)f(y)g(z)dydz
\]
when $f,g \in C^{\infty}({\Bbb T}^d) = C^{\infty}({\Bbb T}_{\hbar
P}^d)$, we expand $f$ and $g$ in Fourier series:
\begin{eqnarray*}
f(x) &= &\sum_m a_me^{2\pi imx} \\
g(x) &= &\sum_n b_ne^{2\pi inx}.
\end{eqnarray*}
Then

\begin{eqnarray*}
(f *_{\hbar} g)(x) &= &\sum_{m,n \in {\Bbb Z}^d} a_mb_ne^{-\pi i\hbar
P(m,n)}e^{2\pi i(m+n)x}
\\
&= &\sum_{m,n \in {\Bbb Z}^d} \int_{{\Bbb T}^d}
e^{-2\pi imy}f(y) dy \int_{{\Bbb T}^d}
e^{-2\pi inz}g(z)dz~e^{-\pi i\hbar P(m,n)}e^{2\pi i(m+n)x} \\
&= &\int_{{\Bbb T}^d \times {\Bbb T}^d}
\sum_{m,n\in {\Bbb Z}^d } e^{-\pi i\hbar P(m,n) - 2\pi i(m(y-x) +
n(z-x))}f(y)g(z)dydz,
\end{eqnarray*}
where all quantities are considered as distributions, so that the
interchange of summation and integration is justified.
Thus, the
kernel of multiplication for our quantum torus ${\Bbb T}_{\hbar P}^d$
is the distribution
\[
K_{\hbar P}(x,y,z) = \sum_{m,n  \in {\Bbb Z}^d}
e^{-\pi i\hbar P(m,n) - 2\pi i(m(y-x) + n(z-x))}.
\]
Being translation invariant, the kernel can be written as $K_{\hbar
P}(x,y,z) = L_{\hbar P}(y-x,z-x)$, where
\begin{equation}
\label{eqn3}
L_{\hbar P}(y,z) = \sum_{m,n \in {\Bbb Z}^d}
e^{-\pi i\hbar P(m,n) - 2\pi i(my + nz)}.
\end{equation}
It is this ``convolution kernel'' $L_{\hbar P}$, defined as a
distribution on the group
${\Bbb T}^d \times {\Bbb T}^d$ by the sum
(\ref{eqn3}), which we will analyze in the rest of this paper.

\section{Differential equation for the kernel}
\label{sec-two}

Differentiating the typical term in (\ref{eqn3}) by $\hbar$ and
writing $P(m,n)$ as a sum $\sum_{j,k=1}^d P_{jk}m_jn_k$, we find
\begin{eqnarray*}
\frac {d}{d\hbar} e^{-\pi i\hbar P(m,n) - 2\pi i(my + nz)} &= &-\pi i
P(m,n)e^{-\pi i\hbar P(m,n) - 2\pi i(my + nz)} \\
&= &-\pi i \sum_{j,k=1}^d P_{jk}m_jn_k e^{-\pi i\hbar P(m,n) - 2\pi
i(my + nz)} \\
&= &-\pi iP \left( \frac{1}{-2\pi i}\frac {\p}{\p y}, \frac{1}{-2\pi
i}\frac {\p}{\p z} \right) e^{-\pi i\hbar P(m,n) - 2\pi i(my + nz)}.
\end{eqnarray*}
Applying this result to each term in (\ref{eqn3}) and adding, we find
the ``Schr\"odinger equation''
\begin{equation}
\label{eqn4}
i \frac {\p}{\p \hbar} L_{\hbar P}(y,z) = -\frac{1}{4\pi }P \left(
\frac {\p}{\p y},
\frac {\p}{\p z} \right) L_{\hbar P}(y,z).
\end{equation}

Although $P$ is skew symmetric with respect to exchange of its two
arguments, the operator
$-\frac{1}{4\pi } P\left( \frac
{\p}{\p y}, \frac {\p}{\p z} \right)$ on
${\Bbb T}^d \times {\Bbb T}^d$ is {\em self adjoint}.  It is the ``quantum
hamiltonian'' corresponding to the classical hamiltonian, quadratic in
the momenta, $p(y,\eta,z,\zeta) = \frac{1}{4\pi }P(\eta,\zeta)$.  If $P$ is
nondegenerate, this hamiltonian is the ``kinetic energy'' and
$-p\left( \frac {\p}{\p y}, \frac {\p}{\p z} \right)$ half the
``laplacian'' for a pseudoriemannian metric of signature $(d,d)$ on
${\Bbb T}^d \times {\Bbb T}^d = {\Bbb T}^{2d}$.  (A similar statement
may be made when $P$ is degenerate, except that the metric is now
``codegenerate''.)

Finally, the ``initial condition'' for $L$, when $\hbar = 0$, is just
the delta function $\delta(y,z)$ at the zero element of the group
${\Bbb T}^d \times {\Bbb T}^d$.

We may summarize the discussion above as follows:
\begin{quote}
\em
The evolution in
$\hbar$ of the convolution kernel of multiplication for the quantum
tori ${\Bbb T}_{\hbar P}^d$ is the same as the time evolution of a
quantum ``free particle'' on the (ordinary) torus ${\Bbb T}^{2d}$ with
indefinite quadratic hamiltonian given by the Poisson structure
$P/2\pi $.
This particle is initially concentrated at the zero element of ${\Bbb
Z}^{2d}$ but becomes completely nonlocalized as soon as $\hbar \ne 0$.
\end{quote}
In other words, quantization appears to be a kind of ``integration''
of a bi-differential operator, the Poisson structure, to a 1-parameter
family (``group'' in some sense?) of Fourier bi-integral operators,
the noncommutative multiplications.

\section{The quantization kernel as a theta function}
\label{sec-three}

The formula (\ref{eqn3}) exhibits the convolution kernel $L_{\hbar
P}$ as a multidimensional theta function.  Here, we should think of
$(y,z)$ as a ``single'' variable in ${\Bbb R}^{2d}$ and $P(m,n)$ as a
quadratic form in the variable $(m,n) \in {\Bbb Z}^{2d}$ (rather than
${\Bbb Z}^d \times {\Bbb Z}^d$).  Since the imaginary part of the
matrix $P(m,n)$ is zero,
we are on the boundary of the region where the function
$\vartheta(\vec{z},\Omega )$ (see page 118 of \cite{mu:tataI})
is holomorphic.

In some special situations, we can diagonalize the quadratic form
$P(m,n)$ by a $2d \times 2d$ matrix which nearly preserves the lattice
$(2\pi{\Bbb Z})^{2d}$.  This allows us to express $L_{\hbar P}(y,z)$
in terms of theta functions of a single variable.

We will confine our attention to the simplest case $d = 2$.  In higher
dimensions, interesting problems in the ``symplectic geometry of
numbers'' should arise; we will not deal with them here.

In coordinates $(x_1,x_2)$ $(\mbox{mod } {\Bbb Z})$ on ${\Bbb T}^2$, the
most general translation-invariant Poisson structure is a constant
multiple of $\frac {\p}{\p x_1} \wedge \frac {\p}{\p x_2}$.  Since we
already have the multiplier $\hbar$ at our disposal, we will assume
that $P = \frac {\p}{\p x_1} \wedge \frac {\p}{\p x_2}$.  Then
${\tilde P}(dx_1) = \frac {\p}{\p x_2}$ and ${\tilde P}(dx_2) =
-\frac {\p}{\p x_1}$, so ${\tilde P}$ is represented with respect to
the standard
bases of tangent and cotangent vectors by the matrix $\pmatrix 0 & -1
\\ 1 & 0 \endpmatrix$.

If we denote the coordinates on ${\Bbb T}^2 \times {\Bbb T}^2$ by
$(y_1,y_2,z_1,z_2)$ and the corresponding coordinates on ${\Bbb
R}^{2*} \times {\Bbb R}^{2*}$ by $(m_1,m_2,n_1,n_2)$, then the
quadratic form $P(m,n)$ is $m_1n_2 - m_2n_1$, and the convolution
kernel for multiplication is
\[
L_{\hbar P}(y,z) = \sum_{(m_1,n_1,m_2,n_2) \in
{\Bbb Z}^4} e^{-\pi i\hbar(m_1n_2 - m_2n_1) -2\pi i(m_1y_1 + m_2y_2 + n_1z_1 +
n_2z_2)}.
\]

To diagonalize the quadratic form $P(m,n)$, we introduce the
coordinates $u_1 = m_1 + n_2$, $v_1 = m_1 - n_2$, $u_2 = m_2 - n_1$,
$v_2 = m_2 + n_1$.  The inverse transformation is $m_1 = \frac {1}{2}
(u_1 + v_1)$, $n_1 = \frac {1}{2} (v_2 - u_2)$, $m_2 = \frac {1}{2}
(u_2 + v_2)$, $n_2 = \frac {1}{2} (u_1 - v_1)$.  Integer values of
$(m_1,m_2,n_1,n_2)$ correspond to integer values of
$(u_1,u_2,v_1,v_2)$ for which $u_1 \pm v_1$ and $u_2 \pm v_2$ are {\em
even}, conditions which define a sublattice $\Lambda \times \Lambda$
of ${\Bbb Z}^2 \times {\Bbb Z}^2$.  Now we have
\newpage
$$L_{\hbar P}(y,z) = \mbox{\hspace{4in}}$$
$$\sum_{(u_1,v_1,u_2,v_2) \in
\Lambda \times \Lambda} e^{(-\pi i\hbar/4) (u_1^2 - v_1^2 +
u_2^2 - v_2^2) -\pi i(u_1(y_1 + z_2) + u_2(y_2 - z_1) +
v_1(y_1 - z_2) + v_2(y_2 + z_1))} =$$
$$\sum_{(u_1,v_1,u_2,v_2) \in \Lambda \times \Lambda}
e^{-\pi i\hbar u_1^2/4 -\pi i u_1(y_1+z_2)}
e^{\pi i\hbar v_1^2/4 -\pi i v_1(y_1-z_2)} \mbox{\hspace{1in}}$$
$$\mbox{\hspace{2in}}\times e^{-\pi i\hbar u_2^2/4 -\pi i u_2(y_2-z_1)}
                            e^{\pi i\hbar v_2^2/4 -\pi i v_2(y_2+z_1)}. $$

We would have succeeded by the
last expression in splitting $L_{\hbar P}(y,z)$ into a product of
single variable theta functions were it not for the fact that the
lattice $\Lambda \times \Lambda$ makes the variables
$(u_1,v_1,u_2,v_2)$ dependent on one another.

Still, things are not so bad, since $(u_j,v_j)$ belongs to $\Lambda$ if
and only if $u_j$ and $v_j$ are either both even or both odd.  Thus,
$\Lambda$ is the union of $2{\Bbb Z} \times 2{\Bbb Z}$ and $(2{\Bbb Z}
+ 1) \times (2{\Bbb Z} + 1)$, and we can write $L_{\hbar P}(y,z)$,
with all sums over ${\Bbb Z}$, as
$$
\left( \sum_n e^{-\pi i\hbar n^2 -2\pi i n(y_1+z_2)}
       \sum_n e^{ \pi i\hbar n^2 -2\pi i n(y_1-z_2)} \right. +  $$
$$    \left.\sum_n e^{-\pi i\hbar (n+\half )^2 -2\pi i (n+\half )(y_1+z_2)}
       \sum_n e^{ \pi i\hbar (n+\half )^2 -2\pi i (n+\half )(y_1-z_2)}
       \right)   \times 	  $$
$$        \left( \sum_n e^{-\pi i\hbar n^2 -2\pi i n(y_2-z_1)}
       \sum_n e^{ \pi i\hbar n^2 -2\pi i n(y_2+z_1)} \right. +   $$
$$        \left.\sum_n e^{-\pi i\hbar (n+\half )^2 -2\pi i (n+\half
        )(y_2-z_1)}
       \sum_n e^{ \pi i\hbar (n+\half )^2 -2\pi i (n+\half )(y_2+z_1)}
       \right)   $$

Using the ``half-integer theta functions'' and some of the simple
theta identities given in Section I.4 of \cite{mu:tataI} (see
especially page 17), we can write:
$$ L_{\hbar P}(y,z)= \mbox{\hspace{4in}}$$
$$ \mbox{\hspace{.4in}}(\vartheta _{00}(y_{1}+z_{2},-\hbar)\vartheta
_{00}(y_{1}-z_{2},\hbar) +\vartheta _{10}(y_{1}+z_{2},-\hbar)\vartheta
_{10}(y_{1}-z_{2},\hbar)) ~~\times $$
$$\mbox{\hspace{.6in}}(\vartheta _{00}(y_{2}-z_{1},-\hbar)\vartheta
_{00}(y_{2}+z_{1},\hbar) +\vartheta _{10}(y_{2}-z_{1},-\hbar)\vartheta
_{10}(y_{2}+z_{1},\hbar)).$$

Alternatively, we can express everything in terms of the basic theta function
$$ \vartheta(z,\tau )=\sum _{n\in {\Bbb Z}} e^{\pi in^{2}\tau +2\pi inz}$$
as:
\newpage
\begin{equation}
\label{eq-kerneltheta}
L_{\hbar P}(y,z)= \mbox{\hspace{4in}}
\end{equation}
$$ (\vartheta(y_{1}+z_{2},-\hbar)\vartheta
(y_{1}-z_{2},\hbar) + e^{-2\pi iy_{1}}\vartheta
(y_{1}+z_{2}+\half
\hbar,-\hbar)\vartheta(y_{1}-z_{2}-\half \hbar,\hbar)) ~~\times $$
$$\mbox{\hspace{.3in}}(\vartheta(y_{2}-z_{1},-\hbar)\vartheta
(y_{2}+z_{1},\hbar) +  e^{-2\pi
iy_{2}}\vartheta(y_{2}-z_{1}+\half
\hbar,-\hbar)\vartheta(y_{2}+z_{1}-\half \hbar,\hbar)).$$

\section{Discussion}
\label{sec-four}

J. J. Duistermaat has pointed out that, although the theta function
(distribution) can be continued from the real axis to a complex half
plane, the presence in (\ref{eq-kerneltheta}) of products in which the theta
function is evaluated at both $\hbar$ and $-\hbar$ shows
that the kernel $L_{\hbar P}$ itself admits no such continuation.  We
hope that a better understanding of the analytic properties of $L$ can
come from its expression in terms of theta functions, not only in the
simple case $d = 2$ considered above, but also in general, where there
may always be a local decomposition similar to (\ref{eq-kerneltheta}).

On the other hand, our work suggests the possibility of something even
more interesting---the application of the extensive $C^*$-algebraic
theory of quantum tori (once again, we cite \cite{ri:tori} for a
survey of this theory) to derive new results, or
reinterpret old ones, about the classical theta functions.
Note that the associativity of the $*_{\hbar}$ product becomes an
(integral) identity for the theta functions.  Is it interesting?  Is
it known?

A symplectic geometric construction of quantum tori was given in
\cite{we:rotation}.  The symplectic groupoid structures on $T^{*}{\Bbb
T}^{d}$ described in that paper may be interpreted as the wavefront
sets (or, more properly, ``frequency sets'', since asymptotics in
$\hbar$ are involved) of the kernels discussed here.

Finally, we return to the Schr\"odinger equation
\[
\left( i \frac {\p}{\p \hbar} + \frac{1}{4\pi }P\left( \frac {\p}{\p y}, \frac
{\p}{\p z} \right) \right) K_{\hbar}(x,y,z) = 0
\]
for which $K$ is the ``fundamental solution'' in the sense that
$K_0(x,y,z) = \delta(y-x,z-x)$.  Do all ``interesting'' quantizations
satisfy an equation of this nature?  (The Moyal quantization of ${\Bbb
R}^d$ does.)  Finding such an equation would be an important step
toward determining the ``natural'' deformation quantizations of given
Poisson structures.

A major obstacle to extending the Schr\"odinger equation to the
general case is that, in the expression $P\left( \frac {\p}{\p y},
\frac {\p}{\p z} \right)$, the vectors $\frac {\p}{\p y}$ and $\frac
{\p}{\p z}$ may live at different points.  There are, however, some
interesting quantizations (notably, the case of quantum groups) of
manifolds with Poisson structures admitting a simple expression in
terms of globally defined vector fields.

\end{document}